\documentclass[aps,prl,twocolumn,groupedaddress]{revtex4}
\usepackage{graphicx}
\usepackage{amsbsy,amssymb,amsmath,bm}
\newcommand{\SRO}{SrRuO$_3$}
\newcommand{\PMO}{PrMnO$_3$}
\newcommand{\CMO}{CaMnO$_3$}
\newcommand{\PCMO}{Pr$_{0.5}$Ca$_{0.5}$MnO$_3$}
\newcommand{\LAO}{LaAlO$_3$}
\newcommand{\STO}{SrTiO$_3$}
\newcommand{\ve}{\varepsilon}
\begin{document}


\title{Tunable magnetic interaction at the atomic scale in oxide heterostructures}

\author{J. W. Seo$^{1,2}$, W. Prellier$^3$, P. Padhan$^{3,4}$, P. Boullay$^3$, J.-Y. Kim$^5$, Hangil Lee$^{5,6}$, C. D. Batista$^{7}$, I. Martin$^{7}$, Elbert. E. M. Chia$^{2}$, T. Wu$^{2}$, B. -G. Cho$^{8}$ and C. Panagopoulos$^{1,2,9}$}
\affiliation{$^1$Cavendish Laboratory, University of Cambridge, Cambridge CB3 0HE, UK}
\affiliation{$^2$Division of Physics and Applied Physics, Nanyang Technological University, 637371 Singapore}
\affiliation{$^3$Laboratoire CRISMAT, CNRS UMR 6508, ENSICAEN, 14050 Caen Cedex, France}
\affiliation{$^4$Department of Physics, Indian Institute of Technology Madras, Chennai - 600036, India}
\affiliation{$^5$Pohang Accelerator Laboratory, Pohang 790-784, South Korea}
\affiliation{$^6$Department of Chemistry, Sookmyung Women's University, Seoul 140-742, South Korea}
\affiliation{$^7$Theoretical Division, Los Alamos National Laboratory, Los Alamos, New Mexico 87545, USA}
\affiliation{$^8$Department of Physics, Pohang University of Science and Technology (POSTECH), Pohang 790-784, South Korea}
\affiliation{$^9$Department of Physics, University of Crete and FORTH, 71003 Heraklion, Greece}


\begin{abstract}
We report on a systematic study of a number of structurally identical
but chemically distinct transition metal oxides in order to determine
how the material-specific properties such as the composition and the
strain affect the properties at the interface of heterostructures. Our
study considers a series of structures containing two layers of
ferromagnetic SrRuO$_3$, with antiferromagnetic insulating manganites
sandwiched in between. The results demonstrate how to control the
strength and relative orientation of interfacial ferromagnetism in
correlated electron materials by means of valence state variation and
substrate-induced strain, respectively.
\end{abstract}

\maketitle

The Giant Magneto-Resistance effect in metallic magnetic multilayer
systems forms the basis of highly successful magnetic sensing and
storage technology \cite{1,2}. The active search for new materials that
would allow for ever higher sensitivity and controllability is under
way. Transition metal oxides (TMO) are particularly attractive, since
there is a plethora of isostructural materials with a wide variety of
magnetic and electronic properties, which can be seamlessly built into
complex heterostructures \cite{3,4,5,6,7,8,9,10}. In heterostructures composed of
different TMO, the disruption introduced even by an ideal interface,
can drastically upset the delicate balance of the competing
interactions among electronic spins, charges and orbitals, leading to
a range of exotic phenomena, including interfacial conduction,
magnetism, and superconductivity \cite{3,4,5,6,7,8,9,10}.  Among TMO, Mn-based
perovskites AMnO$_3$ (manganites) are one of the best studied classes of
materials that exhibit particularly rich set of behaviors tunable by
composition, pressure and temperature \cite{11}. It has been recently demonstrated that
when put into contact with ferromagnetic (FM) \SRO, the interface
layer of antiferromagnetic (AF) manganite becomes FM \cite{Choi-SMOSrRuO}. In this
work we report a strategy and evidence for manipulating the magnetic
properties at the atomic level in digitally synthesized
nano-heterostructures. We engineer interfacial FM
in correlated electron materials by means of valence state variation
and substrate-induced strain.

\begin{figure}[!h]
\includegraphics[scale=0.45]{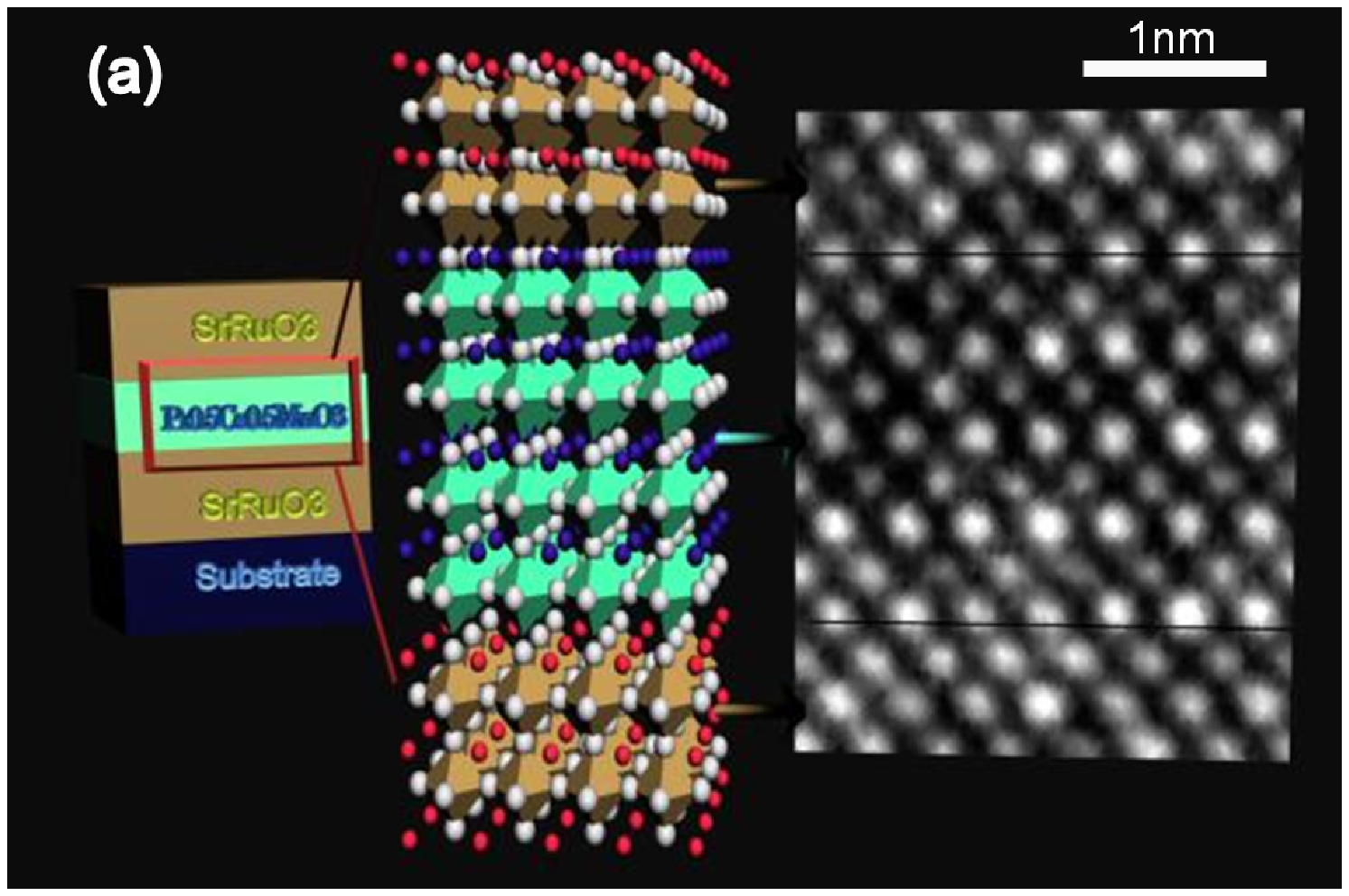}\\
\includegraphics[scale=0.55]{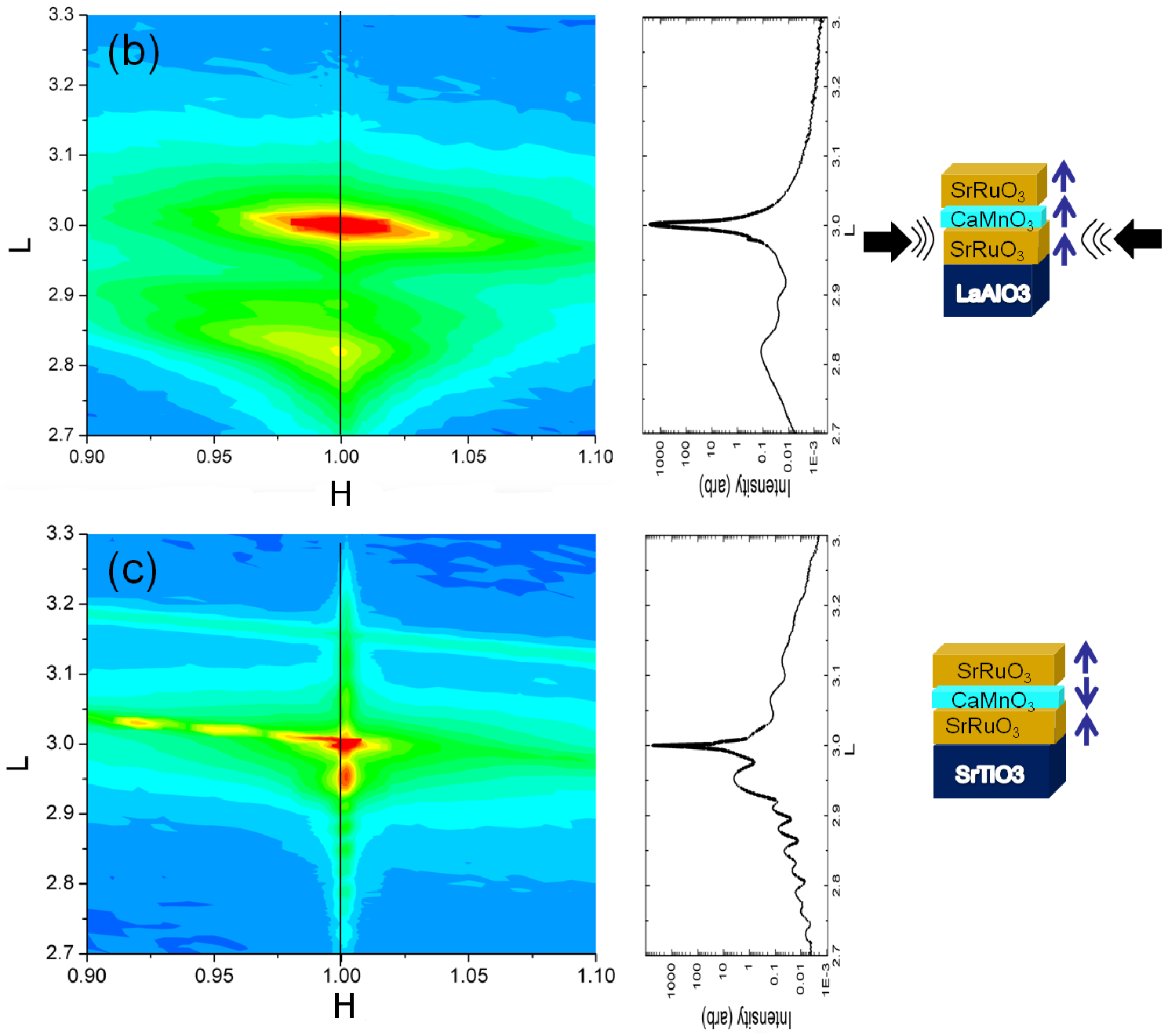}
\caption{\label{fig_structure} (a) Schematic structure (left and middle panels) of the trilayer [18 u.c SrRuO$_{3}$/4 u.c Pr$_{0.5}$Ca$_{0.5}$MnO$_{3}$/18 u.c SrRuO$_{3}$] on SrTiO$_3$. 
The high resolution electron microscope (HREM) image (right panel) shows the epitaxial growth of the sample on SrTiO$_3$. 
Reciprocal space maps (left panels) of the trilayer [SrRuO$_{3}$/CaMnO$_3$/SrRuO$_{3}$] grown on (b) LaAlO$_3$ and (c) SrTiO$_3$ near ($hkl$)=(103). 
The schematic at the right hand side of panel (b) depicts the switching/reversal of the spin direction along the $c$ axis due to the effect of strong compressive strain arising from the LaAlO$_3$ substrate.
}
\end{figure}

Trilayers, [SrRuO$_{3}$/manganite/SrRuO$_{3}$], were grown simultaneously on
(001)-oriented SrTiO$_{3}$ and LaAlO$_{3}$ substrates using the multitarget
pulsed laser deposition technique with energy density of 3$J/cm^{2}$. The conditions for optimizing the deposition can be found elsewhere \cite{12}. In this study, we chose three types of AF insulating manganites: CaMnO$_{3}$ with Mn$^{4+} ({\rm orbital\ state\ }t^{3}_{2g})$, Pr$_{0.5}$Ca$_{0.5}$MnO$_{3}$ with Mn$^{3+/4+}$ and PrMnO$_{3}$ with Mn$^{3+} (t^{3}_{2g}e^{1}_{g})$.  The schematic drawing and high resolution electron microscopy (HREM) image of the trilayers [18~unit cell (u.c.)\ SrRuO$_{3}$/4~u.c.\  Pr$_{0.5}$Ca$_{0.5}$MnO$_3$/18~u.c.\ SrRuO$_{3}$] deposited on SrTiO$_3$ is depicted
in Fig.~1(a).  HREM was performed on cross-section specimens prepared by mechanical polishing followed by ion-milling. The typical HREM image, viewed along the [001] direction (Fig. 1(a)) reveals the epitaxial growth of the trilayers onto the SrTiO$_{3}$ substrate.
Structural characterization of the trilayers with \CMO\ manganite layer was performed using X-ray diffraction (XRD) \cite{13} and XRD mapping with photon energy of 11KeV as shown in Fig. 1(b) and (c). The XRD maps of a reciprocal space in the vicinity of ($hkl$)=(103) reflection were performed at room temperature at beamline 10A of the Pohang Accelerator Laboratory in Korea using four-circle diffractometers. The typical twin structure of the \LAO\ substrate gives rise to the broad peak at (103) \cite{14}. The red and blue colors indicate high and low scattering intensity, respectively. The maps show the in-plane lattices of the substrates and the films match each other without lattice relaxations. The ratio of in-plane to out-of-plane lattice parameters ($a/c$) estimated by the maps is 0.94 (strong compressive strain) in the case of films grown on LaAlO$_3$ (Fig. 1(b)), whereas $a/c \sim$0.98 (weak compressive strain) in the case of SrTiO$_3$ (Fig. 1(c)). The substrate $c$-axis parameters obtained from $\theta$-2$\theta$ scans (the middle panels of (b) and (c)) at the $h$=1 reflections (guided by the black vertical line on the maps) are 3.789{\AA} and 3.906{\AA}, respectively. An earlier study reports that all the heterostructures we study here have the same Curie temperature (T$_C$) i.e., approximately 150K, close to the T$_C$ of the FM SrRuO$_3$ film and independent of the substrates and AF spacers \cite{13}. This small variation in T$_C$ indicates the absence of intermixing at the interfaces.  The coercive fields measured at 80K in a magnetic field applied perpendicular to the films (parallel to the easy axis) are 0.2T and 0.5T for the films grown on SrTiO$_3$ and LaAlO$_3$, respectively. The saturated magnetic moments vary from 1 to 1.3$\mu_B$/unit cell depending on the manganite layers.
\begin{figure}
\includegraphics[scale=0.55]{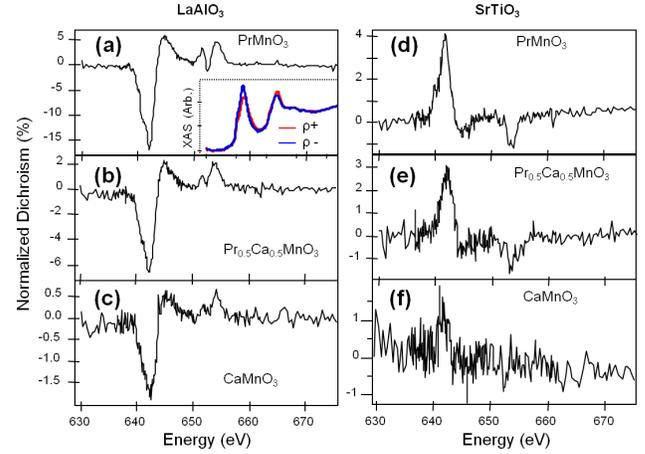}
\caption{\label{fig2} Mn L$_{2,3}$ edge XMCD (black: $\rho^+ - \rho^-$) spectra of the manganite layer in the trilayer
[SrRuO$_{3}$/manganite/SrRuO$_{3}$]. Spectra are taken at 80K in a magnetic field of 0.9T applied
perpendicular to the plane of the film. 
(a)-(c) and (d)-(f) depict spectra
for the samples deposited on LaAlO$_{3}$ and SrTiO$_{3}$, respectively. 
Inset in panel (a): Mn L$_{2,3}$ XAS (red: $\rho^+$, blue: $\rho^-$) spectra.}
\end{figure}

We studied the magnetic profile of the manganite layers in the devices using
atom-selective X-ray magnetic circular
dichroism (XMCD) at the absorption edge
of Mn L$_{2,3}$. XMCD measurements
were performed at beamline 2A of the Pohang Accelerator Laboratory
in Korea.
All XMCD data were obtained after zero-field cooling to 80K. We
measured the total yield signal with an energy resolution of
200 meV in a magnetic field of 0.9T, perpendicular to the plane
of the film. Our approach allows for investigation of a buried layer without spurious
signals from neighbouring layers or the substrate. The measurement was performed by saturating the magnetization of SrRuO$_3$ layers in the $c$-direction by a magnetic field in excess of the coercive field.
Figure 2 depicts the Mn dichroism spectra ($\rho^{+}$-$\rho^{-}$), i.e., the
difference between X-ray absorption spectroscopy (XAS) data
taken with the helicity parallel and antiparallel to the applied
magnetic field. The spectra are normalized to the intensity of
$\rho^{+}$ + $\rho^{-}$ at the L$_3$ peak. Notably, Mn dichroism for the samples deposited on LaAlO$_3$ (Fig.~2(a)-(c)) has the opposite sign to those grown on SrTiO$_3$ (Fig.~2(d)-(f)), implying that the substrate determines the {\em sign} of the magnetic coupling between Mn and Ru across the interface. The schematic drawings in Fig.~1(b) and (c) (right panels) show that the spin direction depends on the substrates the films are grown onto. The large arrows in Fig.~1(b) (right panel) indicate the strong compressive strain is due to the LaAlO$_3$ substrate.

Table~1 summarizes the values of the magnetization for the three
different manganite layers as deduced from the normalized XMCD,
($\rho^+$ - $\rho^-$)/($\rho^+$ + $\rho^-$)(\%). In the case of PrMnO$_3$ grown on the LaAlO$_3$ substrate, the value of 17\% has been estimated to correspond approximately to 1.7$\mu_B$/Mn determined by the sum rule \cite{15}. The negative sign
for the samples grown on SrTiO$_3$ indicates that the spins of the
manganite layer point antiparallel to the applied magnetic field. Notably, the magnetization of the manganite layers
deposited on SrTiO$_3$ is weaker compared to the heterostructures
deposited on LaAlO$_3$. These results indicate that the substrate
influences the strength and the direction of the induced
magnetization in the manganite layer. Our findings demonstrate that one may be able to prepare tailor-made devices with the desired direction and magnitude of the moment in the nanometer-thin FM interface, by depositing the multilayer on a pre-specified substrate. Notably, the observed magnitude of the magnetization follows the order: PrMnO$_{3}$ $>$ Pr$_{0.5}$Ca$_{0.5}$MnO$_{3}$ $>$
CaMnO$_{3}$, particularly evident for the case of \LAO\ substrate.

\vspace*{0.5cm}
\begin{table}
\caption{ Magnetization of three types of spacer manganite layers
(CaMnO$_{3}$, Pr$_{0.5}$Ca$_{0.5}$MnO$_{3}$, and PrMnO$_{3}$)
deposited on two different substrates, namely LaAlO$_3$ and SrTiO$_3$. The values of the magnetization are deduced by normalized XMCD signals defined as ($\rho^{+} - \rho^{-}$)/($\rho^{+} + \rho^{-}$). 
}
\vspace*{0.5cm}
\begin{tabular}{|l|c|c|r|}
\hline
&CaMnO$_{3}$ &Pr$_{0.5}$Ca$_{0.5}$MnO$_{3}$ & PrMnO$_{3}$ \\
\hline
LaAlO$_{3}$&    2\% &  6.5\%   & 17\% \\\hline
SrTiO$_{3}$&    -1.5\%&  -3\%&  -4\%\\\hline
\end{tabular}
\end{table}

The sign and strength of the inter-ion magnetic exchange interaction is commonly analyzed within the Goodenough-Kanamori (GK) rules \cite{16}.  Depending on the orbital occupancy and the presence or absence of an overlap between the orbitals belonging to the two ions, the exchange can be either FM or AF.  For instance, for half-filled orbitals with non-zero overlap the exchange is AF \cite{17}. Let us examine what would be the consequences of the GK rules in our system.  Ru$^{4+}$ has four electrons in the $t_{2g}$ orbitals. Three of them are ferromagnetically aligned by the onsite Hund's interaction, while the fourth is antialigned with the first three, yielding a state with spin $S=1$ (the crystal field splitting between $e_g$ and $t_{2g}$ orbitals of Ru exceeds the onsite FM Hund's coupling, which leaves the Ru $e_g$ orbital unoccupied).

In \CMO, Mn state is $t_{2g}^3$ ($S = 3/2$), which corresponds to half-filled $t_{2g}$ band. The Mn atom is connected with Ru via apical oxygen, which leads to one-to-one hybridization between the respective orbitals, e.g. Ru $t_{2g}$ $xz$ ($yz$) with Mn $t_{2g}$ $xz$ ($yz$) and Ru $e_g$ $3z^2 - r^2$ with Mn $e_g$ $3z^2 - r^2$ ($z ||$ interface normal). Note that by symmetry, there is no oxygen-mediated hybridization between the $xy$ orbitals of Ru and Mn (we neglect the possible but small symmetry-breaking effects of lattice distortions), while the direct overlap between the orbitals is very weak. The same is true for
the $x^2-y^2$ $e_g$ orbitals. If Ru were in $t_{2g}^3$ fully polarized state, the GK rules would imply that the coupling between Ru and Mn is AF.
However, experimentally we find that on \LAO\ substrate the coupling is FM. Clearly the FM coupling can only be caused by the fourth -- the $minority$ -- electron of Ru. In the limit of small hybridization between Ru and Mn ions, the sign and magnitude of the magnetic coupling can be understood by means of perturbation theory in hopping $t$. The virtual electron hopping processes transfer electrons between low and high-energy ionic sates (the energy difference must be much bigger than $t$ for the perturbation theory to be valid). FM coupling induced by hopping of the minority electron can become dominant if the energy barrier associated with the transfer of this electron between FM aligned Ru to Mn ions is considerably smaller than for the same process for AF alignment of Ru and Mn. We show in the following that there is indeed a significant range of realistic parameters where FM coupling dominates. Whether or not the minority electron contributes to the magnetic coupling between Ru and Mn is determined by the $strain$, which controls the relative energies of the $t_{2g}$ orbitals of Ru. For strong compressive inplane strain, such as the one induced by the \LAO\ substrate, the $xz$, $yz$ orbitals are lower than the $xy$ orbital (similarly, $e_g$  $3z^2 - r^2$ is lower than $x^2 - y^2$).  These are the orbitals that hybridize across the interface, and therefore when the minority-spin electron of Ru occupies either one of them, it can mediate FM coupling between Ru and Mn. For the tensile strain, however, the minority electron of Ru occupies the $xy$ orbital and does not hybridize with Mn, which leaves only the AF channel open.

To test this qualitative argument, we performed exact diagonalization studies of the Mn$^{4+}$-Ru$^{4+}$ complex including all $d$ orbitals for various values of effective Ru-Mn hopping integral $t$ and strain-induced crystal field splitting,  $\delta = \ve_{xy} - \ve_{xz,yz}$ (Fig.~3).  The hopping integral is taken to be zero between the $d$ orbitals for which by symmetry there is no oxygen-mediated hybridization ($xy$ case). The spin state of isolated Mn ion is $S_{\rm Mn} = 3/2$ and of isolated Ru ion is $S_{\rm Ru} = 1$. The Hamiltonian is
\begin{eqnarray}H_0 &=&   \sum_{j,\alpha} \varepsilon_j n_{j \alpha} + \sum_{j,\alpha} U_j n_{j \alpha \uparrow} n_{j \alpha \downarrow}
+  \sum_{j, \alpha \neq \alpha'} \frac{U_j-J_j}{2} n_{j \alpha} n_{j \alpha'}
\nonumber \\
&-&   \sum_{j, \alpha \neq \alpha'} \frac{J_j}{2} [{\bf S}_{j \alpha} \cdot {\bf S}_{j \alpha'} + \frac{1}{4} n_{j \alpha} n_{j \alpha'} -
  d^{\dagger}_{j\alpha\uparrow} d^{\dagger}_{j\alpha\downarrow}
d^{\;}_{j\alpha'\downarrow} d^{\;}_{j\alpha'\uparrow}]
\nonumber \\
&+&  \delta \sum_{j} (n_{jxz} + n_{jyz} )
+
\sum_{\alpha, \alpha', \sigma} t_{\alpha \alpha'}
(d^{\dagger}_{1\alpha\sigma} d^{\;}_{2\alpha' \sigma} + {\rm H. c.})
\end{eqnarray}
where $j=$Mn$^{4+}$, Ru$^{4+}$, $U_j$ are the Coulomb repulsions and $J_j$ are the Hund's coupling constants.
The orbital label $\alpha$ takes the values $\alpha = \{xy, xz, yz\}$, while $\sigma= \uparrow, \downarrow$ is the spin label.
Finally, $n_{j \alpha \sigma}= d^{\dagger}_{j\alpha\sigma} d^{\;}_{j\alpha\sigma}$ and $n_{j \alpha} = \sum_{\sigma} n_{j \alpha \sigma}$.
The hopping matrix is diagonal, $t_{\alpha, \alpha'} = t_{\alpha, \alpha} \delta_{\alpha,\alpha'}$, with
$t_{xy,xy}=0$, $t_{xz,xz}=t_{yz,yz}=t$. We use parameters \cite{params} $J_{\rm Ru} = J_{\rm Mn}= 1.5$ eV, $U_{\rm Mn} = 5$ eV,  $U_{\rm Ru} = 4$ eV, $ \varepsilon_{\rm Mn} - \varepsilon_{\rm Ru} = -2$ eV. We find that for $\delta \gtrsim \delta_c(t)$ (strong compressive strain) where $\delta_c(t)$ is a critical value of $\delta$, the sates with high
total spin ($S_{tot}=5/2$ and $S_{tot}=3/2$), are stabilized, which corresponds to FM coupling between Mn and Ru. The critical value of $\delta$ for the transition between $S_{tot}=1/2$ and  $S_{tot}=3/2$ depends on the value of $t$ (see Fig.3 (a)). We point out that the state with $S_{tot} = 3/2$ that appears at larger values of hopping $t$ is beyond the ionic picture. On the other hand, for $\delta < \delta_c(t)$ (tensile or weak compressive strain) the Ru and Mn ions are anti-aligned into the state with lowest total spin ($S_{tot}=1/2$).  This is in agreement with the above qualitative discussion.

In \PMO,   the nominal valence of Mn is 3+, i.e., there is one extra electron in the  $e_g$ orbital of Mn. Due to the Hund coupling, the spin of this electron is aligned with the spins of the other ($t_{2g}$) electrons of Mn.
In the case of strong compressive inplane strain (\LAO\ substrate) the extra $e_g$ electron occupies the $3z^2 - r^2$ orbital of Mn, which is well hybridized with the corresponding empty orbital of Ru, and therefore according to the GK rules favors FM alignment of Mn and Ru spins.
This contribution adds to the FM coupling that is already present in the case of \CMO/\SRO\ interface. In addition, the magnetic moment is bigger for the Mn$^{3+}$ ion ($S_{\rm Mn}=2$). Both effects lead to the relatively enhanced value of the induced FM  moment in \PCMO\ and \PMO, in agreement with the progression that we observe experimentally (Table 1).
On the other hand, for tensile or weak compressive inplane strain (\STO\ substrate), the extra electron on Mn occupies the $e_g$ $x^2 - y^2$ orbital, which by symmetry cannot hybridize with the apical oxygen orbitals and thus does not have significant overlap with the Ru d-orbitals across the interface. Thus we conclude that for tensile or weak compressive inplane strain, just as in the case of \CMO, in \PCMO\  and \PMO, there is no virtual hopping process that would favor FM alignment of Mn and Ru ions, and the result is a (relatively weak) AF coupling between Mn and Ru ions.  This is in complete qualitative agreement with our experimental observations.

\begin{figure}
\includegraphics[scale=0.5]{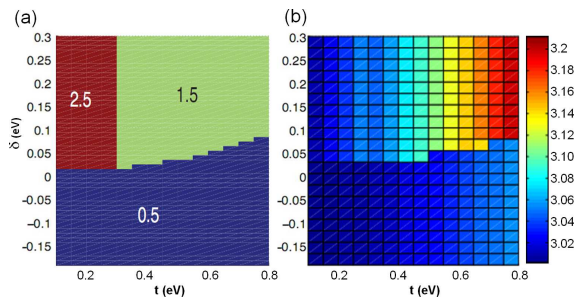}
\caption{\label{ED} (a) The total spin of coupled  Mn${^{4+}}$ ($S = 3/2$) and Ru$^{4+}$ ($S = 1$) as a function of inter-ion $t_{2g}$ hybridization $t$ and the crystal field splitting $\delta = \ve_{xy} - \ve_{xz,yz}$.
(b) The charge of the Mn ion. The transitions between different total spin states are accompanied by the change in the charge of the ions. 
}
\end{figure}
In summary, we report direct evidence for tunable FM behavior at the atomic scale in strongly correlated oxide heterostructures. We find that the orientation and strength of the induced interfacial magnetism can be very sensitive to strain. 
By selecting appropriate substrate one may now design complex magnetic heterostructures with the desired relative arrangement of the magnetic elements, of potential utility to electronics and spintronics, including magnetic memory and sensing. Moreover, the strain can be also induced by means of  externally applied force, which can thus cause {\em mechanically-induced} magnetic reorientation. For example, by using a piezoelectric substrate such as PNM-PT (Pb(Mg$_{1/3}$Nb$_{2/3}$)O$_3$-PbTiO$_3$) \cite{18} one would be able to control the orientation and strength of the magnetization by tuning the lattice parameters by means of an applied electric field.

This work was supported by MEXT-CT-2006-039047, EURYI, KRF-2005-215-C00040, LAFICS, STAR, CEFIPRA/IFPCAR,
National Research Foundation of Singapore, and the Merlion project (n2.04.07).
The authors acknowledge H. Hwang for helpful comments on the Goodenough-Kanamori rules, and A. Pautrat and Y. Tokura for useful discussions.







\

\end{document}